\documentclass[aps,prl,twocolumn,showpacs,superscriptaddress,groupedaddress]{revtex4}

\usepackage{graphicx}  
\usepackage{dcolumn}    
\usepackage{bm}        
\usepackage{amssymb}   

\begin{document}

\title{Coulomb blockade and Bloch oscillations in superconducting Ti nanowires}

\author{J. S. Lehtinen} 
\affiliation{University of Jyv\"{a}skyl\"{a}, Department of Physics, PB 35, 40014 Jyv\"{a}skyl\"{a}, Finland} 

\author{K. Zakharov} 
\affiliation{Physics Faculty, Moscow State University, 119991 Moscow, Russia} 

\author{K. Yu. Arutyunov}
\email{Konstantin.Arutyunov@phys.jyu.fi} 
\affiliation{University of Jyv\"{a}skyl\"{a}, Department of Physics, PB 35, 40014 Jyv\"{a}skyl \"{a}, Finland} 
\affiliation{Nuclear Physics Institute, Moscow State University, 119992 Moscow, Russia} 

\date{\today}

\begin{abstract} 

Quantum fluctuations in quasi-one-dimensional
superconducting channels leading to spontaneous changes of the phase of the
order parameter by $2\pi$, alternatively called quantum phase slips
(QPS), manifest themselves as the finite resistance well below the critical
temperature of thin superconducting nanowires and the suppression of
persistent currents in tiny superconducting nanorings. Here we report the
experimental evidence that in a current-biased superconducting nanowire the
same QPS process is responsible for the insulating state -- the Coulomb
blockade. When exposed to RF radiation, the internal Bloch oscillations can be
synchronized with the external RF drive leading to formation of quantized
current steps on the I-V characteristic. The effects originate from the
fundamental quantum duality of a Josephson junction and a superconducting
nanowire governed by QPS -- the QPS junction (QPSJ).

\end{abstract}



\maketitle

Since the early years of superconductivity zero resistivity and perfect
diamagnetism were considered as the mandatory attributes of a superconducting
state. Later it became clear that in sufficiently small systems thermodynamic
fluctuations of the order parameter may significantly broaden the
superconducting phase transition. In particular case of quasi-one-dimensional
(1D) superconductors \cite{Arutyunov-PhysRep2008} quantum
fluctuations, also called \textit{quantum phase slips} (QPS),
enable finite resistivity in nanowires \cite{Giordaano-PRL88-QPS}, \cite{Bezryadin-Tinkham-Nature2000-QPS}, 
\cite{Tian-PRL2005-QPS}, \cite{Altomare-PRL05-QPS}, \cite{Zgirski-NanoLett05-QPS},
\cite{Zgirski-PRB08-QPS}, \cite{Arutyunov-QPS-in-Ti} and suppress persistent
currents in tiny nanorings \cite{Arutyunov-QPSring-SciReports2012}, 
\cite{Astafiev-coherentQPS-NATURE2012} at temperatures well below
the critical point. Quite recently it has been realized that a superconducting
nanowire governed by quantum fluctuations is dual to a Josephson junction
(JJ): the Hamiltonians describing the two systems are identical with respect
to parametric substitution $E_{QPS} \leftrightarrow E_{J}$,
$E_{L} \leftrightarrow E_{C}$, $I \leftrightarrow V/R_{Q}$ and $q \leftrightarrow \varphi$ \cite{Mooij-Nazarov-NatPhys06}. 
Hence, the extensively developed physics, describing the behavior of a JJ, should be
straightforwardly applicable to such nanowire - the \textit{quantum phase slip junction} (QPSJ).
In the particular case of a current-biased QPSJ, it should exhibit the coherent charge oscillations
qualitatively described by expressions similar to those of Bloch electrons in
periodic potential of a crystal lattice. The experimental test of this
prediction is the main objective of the paper.

\begin{figure} 
	\includegraphics[width=\linewidth]{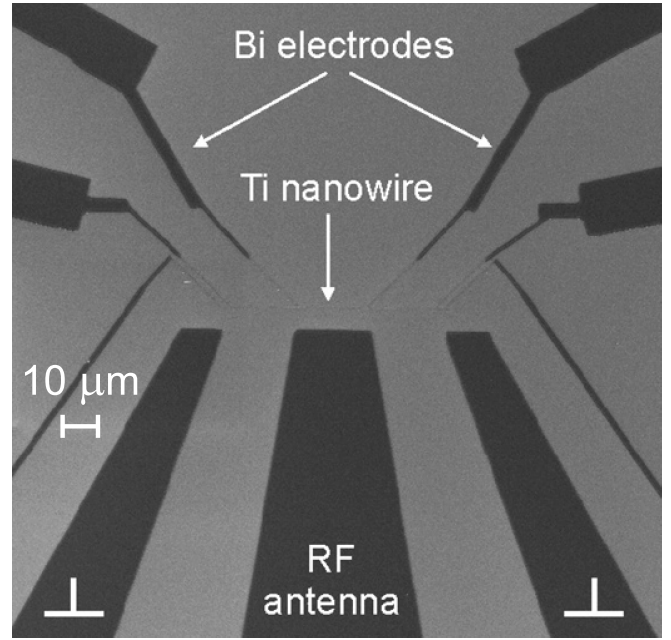}
	\caption{Scanning electron microscope image of a typical sample with high-Ohmic contacts enabling four-probe transport
	measurements and introduction of RF radiation. }
	\label{Fig.1} 
\end{figure}

A conventional Josephson effect is observed in systems with coupling energy
$E_{J}=(R_{Q} / R_{N})(\Delta/2)\gg E_{C}$
and conductance $G \gg 1/R_{Q}$, where $R_{Q} = 6.45 ~k\Omega$,
$R_{N}$ is the junction normal state resistance and $\Delta$
is the superconducting gap. In this limit the superconducting phase
$\varphi$ behaves as a classical variable. Application of
external RF radiation with frequency $f_{RF}$ leads to formation of
quantized voltage steps on the $I-V$ characteristic - Shapiro effect:
$V_{n}=h \times (f_{RF}/2e) \times n$,
$n=0,1,2,\ldots$ In the opposite limit,
$E_{J} \ll E_{C}$ and $G \ll 1/R_{Q}$, the
quasicharge $q$ rather than $\varphi$ is the classical
quantity, and the Coulomb effects take over the Josephson coupling
\cite{Averin-Zorin-Likharev-JETP85}, \cite{Likharev-Zorin-JLTP85}. 
The experimental observation of the charge phenomena in JJs requires two conditions. 
First, the JJ capacitance $C$ should be small providing high charging energy 
$E_{C}=(2e)^{2}/2C \gg E_{J}$. Second, to enable the quasicharge $q$ be
a well-defined quantity, the system should be current biased. The periodic
charging/discharging of the junction leads to Bloch-type oscillations
manifesting as peculiar back-bended $I-V$ characteristic. External RF
radiation can be synchronized with the internal charge oscillations leading to
singularities at quantized values of current $I_{n}=2e \times
f_{RF} \times n$. So far only rather broad $n=1$ singularities
have been reported in ultra-small JJs \cite{Kuzmin-Haviland-PRL91}. Indirectly, the presence of Bloch oscillations
has been demonstrated in another Josephson device - Cooper pair box, where the
injection of current $I$ through the capacitively coupled gate resulted in
formation of narrow side bands $f_{B}=I/2e$ in the spectrum
of the reflected RF signal \cite{Nguyen-PRL07}.

Observation of the dual effect - the charge phenomena in a QPSJ - requires
sufficiently high rate of quantum fluctuations $E_{QPS}$ exceeding the
energy $E_{L}=\Phi_{0}^{2}/2L$, associated with
the system inductance $L$. Here the superconducting flux quantum
$\Phi_{0}= 2.07 \times 10^{-15} ~ Wb$,
$E_{QPS}=\Delta(R_{Q}/R_{N})(L/\xi)^{2} \exp(-S_{QPS})$, $S_{QPS}=A(R_{Q}/R_{N})(L/\xi)$, the numerical factor $A \simeq 1$,
$R_{N}$ is the normal state resistance of the wire with length $L$, and
$\xi$ is the superconducting coherence length \cite{Zaikin-PRL1997-QPS}, \cite{Zaikin-Uspexi98-QPS}. If
$S_{QPS} \gg 1$ the rate of QPSs is small and the nanowire
exhibits almost conventional superconducting properties: vanishingly small
resistance below the critical current $I_{c}$. In the opposite limit
$S_{QPS} \simeq 1$ the quantum fluctuations are strong and,
being current biased, such a (superconducting!) nanowire below the certain
critical voltage $\delta V_{CB}$ should demonstrate the
insulating state - the Coulomb blockade. Note that the model \cite{Zaikin-PRL1997-QPS}, \cite{Zaikin-Uspexi98-QPS} describes the impact of rather weak quantum fluctuations: i.e.
$S_{QPS} \gg 1 $, and has been proven to be in a good
quantitative agreement with experiments \cite{Zgirski-PRB08-QPS}, \cite{Arutyunov-QPS-in-Ti}. In the opposite limit $S_{QPS} \simeq 1$, which is of primary interest for the present paper,
strictly speaking, utilization of the model \cite{Zaikin-PRL1997-QPS}, \cite{Zaikin-Uspexi98-QPS} requires further justification.

The presence of Coulomb blockade in $NbSi$ nanowires has been reported
\cite{VanDerSar-MSc2007}, \cite{Hongisto-QPS-SET-2012}. However in this extreemly high-Ohmic and
strongly disordered superconductor the presence of weak links forming a chain
of JJs cannot be ruled out completely. In our work we study titanium
nanostructures, where it has been already demonstrated that below the
effective diameter $\sigma^{1/2} \simeq 50 ~nm$
the rate of QPSs is sufficiently high to broaden the $R(T)$ phase transition
in nanowires (with low-Ohmic environment) \cite{Arutyunov-QPS-in-Ti} and to suppress the persistent currents in
nanorings \cite{Arutyunov-QPSring-SciReports2012}. The samples
were fabricated using the same technique described in our earlier papers
\cite{Savolainen-APA2004-IBE}, \cite{Zgirski-Nanotechnology2008-IBE}, \cite{Zgirski-PRB08-QPS}, \cite{Arutyunov-QPS-in-Ti}. High-Ohmic probes were fabricated
either from slowly evaporated at high angle titanium, showing no traces of
superconductivity down to 20 mK, or - from bismuth (Fig. 1) with resistance up
to $R_{p} \sim 50 ~M\Omega$ enabling reliable current biasing of the
titanium QPSJ. The extensive microscopic (SEM, TEM and AFM) and elemental
(TOF-ERDA) analyses \cite{Arutyunov-QPS-in-Ti}, \cite{Arutyunov-QPSring-SciReports2012} revealed rather conventional
polycrystalline structure of the samples without obvious structural defects and with the surface roughness $\pm ~2 ~ nm$. The presence of an extended network of weak links, blocking the metal-to-metal electric current, looks rather unlikely.

At a given temperature $T < T_{c}$ the relatively thick samples with
diameter $\sigma^{1/2} \simeq 40 ~nm$ and
low-Ohmic probes with $R_{p} \simeq 15 ~k\Omega$ indeed demonstrate
weak Coulomb blockade with the zero-bias conductivity lower than at a finite
bias, but higher than in the normal state (Fig. 2). The observation indicates
that in these samples with parameter $S_{QPS} \simeq 14$ and the
associated energy $E_{QPS} \simeq 0.1 ~\mu eV$ the
residual superconductivity `wins' over the charge effects poorly resolved at
realistically obtainable temperatures. Application of external RF radiation
stimulates weak non-linearities on the $dV/dI$ dependencies at currents
$I_{n}=2e \times f_{RF} \times n$ (Fig. 2). Note
that the corresponding positions in voltage scale $V_{n}(f_{RF})$ do
not form any rational Shapiro pattern (Fig. 2, inset) to be present in a
conventional Josephson system.
\begin{figure}
	\includegraphics[width=\linewidth] {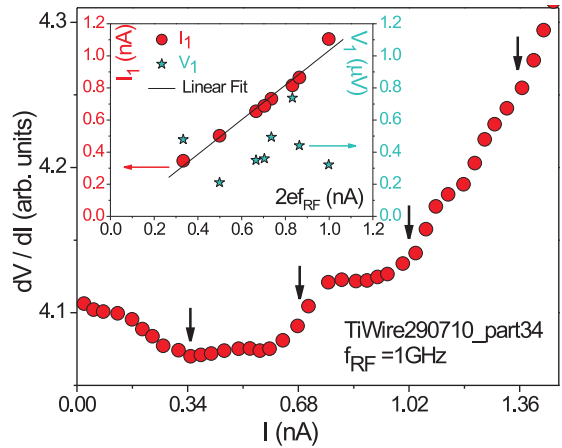}
 	\caption{(Color online). All-titanium structure: the nanowire length $L=20 ~\mu m$, 
	$\sigma^{1/2}=40 \pm 2 ~nm$ and $R_{p} \simeq 15 ~k\Omega$. $dV/dI(I)$ in presence of external RF radiation with
	frequency $f_{RF}=1 ~GHz$ at $T=70~mK$. Arrows indicate the positions of the
	expected current singularities $I_{n}=2e \times f_{RF}\times n$. Inset: positions of the first current
	singularity $I_{1}$ (left axis, circles) and the corresponding voltage
	$V_{1}$ (right axis, stars) as function of the RF frequency
	$f_{RF}$. Note the acceptable linear fit (solid line) for the current
	singularities, while absence of any rationality for the $V_{1}(f_{RF})$ dependency, which one might expect in case of a conventional
	Shapiro effect.}
	\label{Fig.2}
\end{figure}
Thinner $\sigma^{1/2} \simeq 24 ~nm$ samples
with relatively high-Ohmic $R_{p} \simeq 10 ~M\Omega$ probes exhibit clear Coulomb blockade with the gap 
$\delta V_{CB} \simeq 0.6 ~mV$ (Fig. 3a) corresponding to the
estimation $\delta V_{CB} \simeq E_{QPS}$ with
parameters deduced from the earlier experiments on similar titanium
nanowires \cite{Arutyunov-QPS-in-Ti}. At small bias currents $I$
$\leq 50 ~pA$ the $V-I$ characteristics of all three samples
demonstrate discontinuity-type switching from the current-carrying to the
insulating state. Such behavior is expected due to the non-single-valued I-V
dependency \cite{Averin-Zorin-Likharev-JETP85}, \cite{Likharev-Zorin-JLTP85}.
The Coulomb gap can be quasi-periodically modulated by the gate potential 
$\delta V_{CB}(V_{GATE})$ with the `RF antenna' electrode used as a DC gate (Fig. 3a, left inset). The
period of the gate modulation is in a reasonable agreement with the geometry
of the experiment resulting in the highest amplitude and the smallest period
for the closest to the gate Sample \#23, and - larger period for
the remote Sample \#1. The Coulomb gap decreases with temperature
and disappears at the critical temperature of bulk titanium (Fig. 3a, right
inset). At a given (low) temperature the Coulomb gap can be eliminated by
application of sufficiently strong magnetic field. External radiation with
frequency $f_{RF}$ generates peculiarities on I-V characteristics at
positions $I_{n}=2e \times f_{RF} \times n$ (Fig.
3b). Electrodynamics of a JJ and a QPSJ is qualitatively indistinguishable.
Association of our results with one of those two systems requires extra
information. However, the explanation based on a conventional single electron
transitor (SET) effect, due to accidental formation of tunnel junction(s) in
the titanium nanowires, most likely, is not credible. First, the microscopic
and elemental analyses do not reveal any presence of weak links \cite{Arutyunov-QPS-in-Ti}, 
\cite{Arutyunov-QPSring-SciReports2012}. Second, the effective capacitance
$C_{eff}$, defining the charging energy $(2e)^{2}/(2C_{eff})=\delta V_{CB}$, is about two orders larger
than of a hypothetical parallel-plate capacitor to be formed in a break of a
nanowire with $\sim 20 ~nm$ diameter. Third, if nevertheless
unintentionally the junctions were formed, it would be very unrealistic that
they provide basically the same charging effect in each of the three samples:
$\delta V_{CB}=0.6, ~0.7$ and $0.8 ~ mV$, respectively (Fig.3a).
And forth, the Coulomb gap disappears above the critical temperature and
magnetic field for titanium. The observation does not support the
interpretation based on existence of rogue tunnel junctions, which would
otherwise enable some residual Coulomb effects in the normal state. Note that
the observed Coulomb gap $2e \times \delta V_{CB}/k_{B}$ corresponds to temperature of several Kelvin, which is
order of magnitude higher than the temperature where the last traces of the
Coulomb gap disappear (Fig. 3a, right inset). Summarizing, in both types of
structures (Figs. 2 and 3) the existence of unintentionally formed tunnel
junction(s) is highly improbable, and the charge phenomena most likely
originate from the QPSs providing the `dynamically driven' equivalent of a JJ
- the QPSJ.
\begin{figure}
	\includegraphics[width=\linewidth]{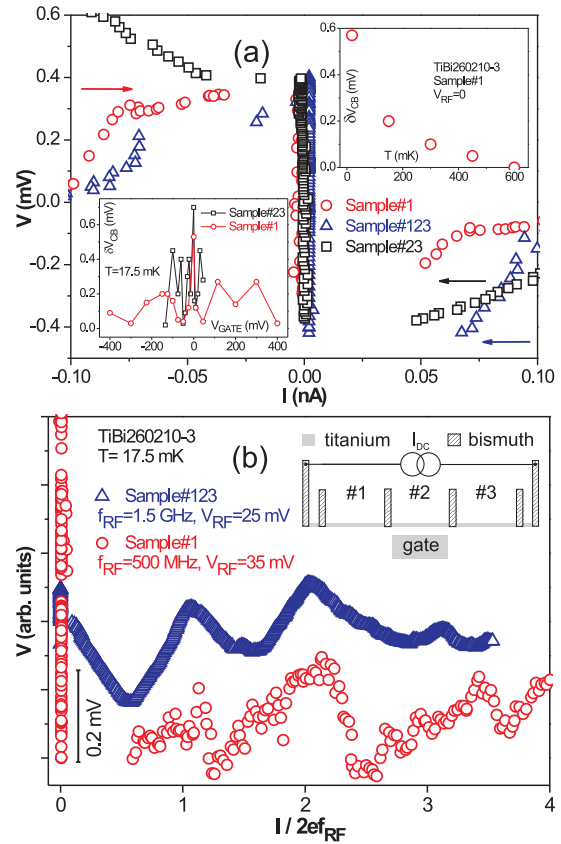}
	\caption{(Color online). Multiterminal titanium nanostructure with three adjacent nanowires each with length 
	$L=20	~\mu m$, $\sigma^{1/2}= 24 \pm 2 ~nm$ and $R_{p} \sim10 ~ M\Omega$ bismuth contacts. (a) The
	$I-V$s demonstrate the Coulomb blockade for all three neighboring parts of
	the same nanostructure. Arrows indicate the direction of the current sweep.
	The Coulomb gap $\delta V_{CB}$ decreases with increase of the
	temperature and disappears above $\sim 450 ~mK$ (right inset).
	Application of the gate voltage $V_{GATE}$ quasi-periodically modulates
	the Coulomb gap (left inset). (b) Application of external radiation with the
	frequency $f_{RF}$ generates non-monotonous peculiarities at positions
	$I_{n}=2e\times f_{RF} \times n$. Inset:
	schematics of the structure.}
	\label{Fig.3}
\end{figure}
The thinnest nanowires with the effective diameter $\sigma^{1/2} \lesssim 18 ~nm$ and the parameter $S_{QPS} \simeq 1$ demonstrate very pronounced back-bended $I-V$
characteristic with the RF induced singularities up to $n = 8$  (Fig 4).
However, the large value of the Coulomb gap (Fig. 4a, left inset), which does
not disappear above the $T_{c}$ of superconducting titanium, leads to a
conclusion that, though unintentionally, some weak links were formed. Note
that all RF-induced singularities disappear above 200 mK, while the
size-dependent critical temperature of the thinnest samples is expected to be
below 250 mK \cite{Arutyunov-QPS-in-Ti}. At higher frequencies
$f_{RF}$ the rich structure develops at the $I-V$ dependencies (Fig.
4b). The positions of the singularities form the regular pattern (Fig. 4c):
$I(n,m)=e \times (n/m) \times f_{RF}$, where the
principal steps ($m = 1$) with $n=1,2,3,4 \ldots$ can
be associated with single electron transport, while the even ones with
$n =2,4,6 \ldots $ - also with Cooper pairs. First
single electron sub-harmonic $n = 1$ and $m = 2$ (step $1/2$) is also resolved.
Co-existence of superconducting and single electron Bloch steps has been
earlier reported in JJs, though only for $n =1$ \cite{Kuzmin-Pashkin-PhysicaB94}.
Remarkably the dependence of the step width $\delta V_{n}$ on amplitude of the RF signal $V_{RF}$ is
essentially non-monotonous (Fig. 4b, inset) following the theoretical
prediction $\delta V_{n} \sim(-1)^{n}J_{n}(V_{RF})$, where $J_{n}$ is Bessel function \cite{Averin-Odintsov-FNT90-16}.
\begin{figure}
	\includegraphics[width=\linewidth]{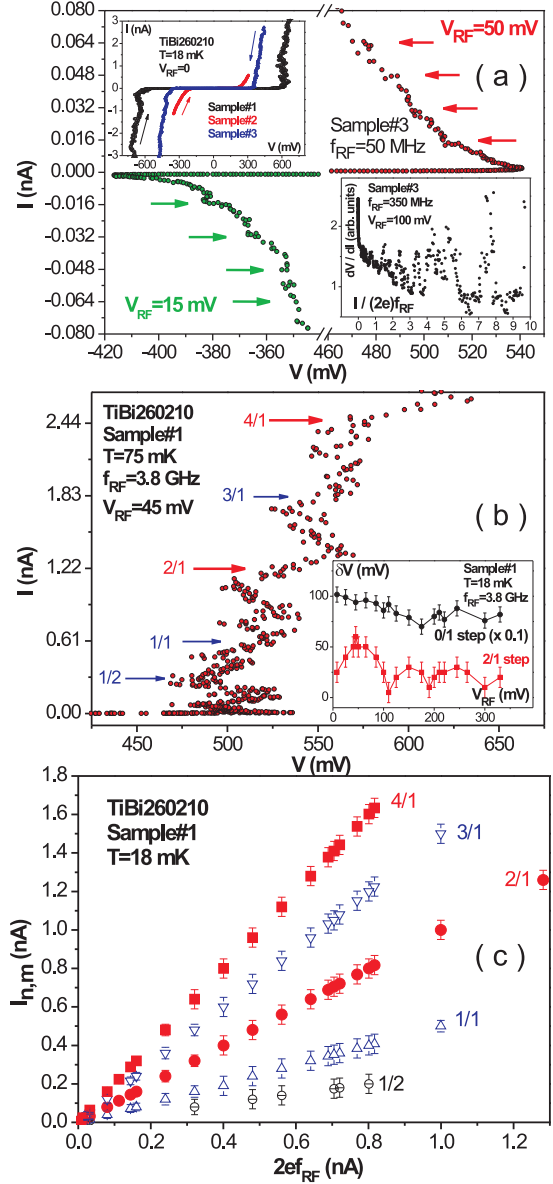}
	\caption{(Color online). Multiterminal nanostructure with each nanowire length $L=20 ~\mu m$ and the
	high-Ohmic bismuth contacts. (a) Sample \#3 with $\sigma^{1/2} =15 \pm 3 ~nm$ and $R_{N} \approx 1.5 ~M\Omega$.
	Zoom of the $I-V$ characteristic at the transition point
	from insulating to current-carrying state at $f_{RF} = 50 ~MHz$ and two
	amplitudes $V_{RF}$ of the RF signal. Arrows indicate the positions of
	the expected current singularities 
	$I_{n}=2e\times f_{RF} \times n$. Note the characteristic back-bended shape of
	the $I-V$ dependence: the `\textit{Bloch nose}'. Left inset:
	larger scale $I-V$ characteristics of the three adjacent sections of the
	same nanostructure. Arrows indicate the directions of the current sweep. Right
	inset: $dV/dI(I)$ at $f_{RF}=350 ~MHz$. (b) Sample \#1 with
	$\sigma^{1/2} =12 \pm 4 ~nm$ and
	$R_{N} \approx 2.5 ~M\Omega$. Magnified view of the current
	singularities at $f_{RF}=3.8 ~GHz$. Arrows indicate the positions of the
	expected current steps. Inset: dependence of the step width $\delta V$ 
	on RF magnitude for the first Cooper pair singularity (2/1) and the
	Coulomb gap (0/1). Note that for the Coulomb gap the scale is reduced by
	factor of 10. (c) Positions of the current singularities $I_{n,m}$ as
	functions of the RF frequency $f_{RF}$.}
	\label{Fig. 4}
\end{figure}
Interpretation of the thinnest sample data (Fig. 4) is less straightforward.
The nanowire sheet resistivity $R_{\square}$ from $0.4$ to $1.9 ~
k\Omega$ is still on the metallic side of the metal-to-insulator transition.
Coulomb effects in titanium have been observed so far in deliberately oxidized
nanowires with noticeably higher resistivity \cite{Schollmann-APL2000}, \cite{Johannson-Haviland-PhysicaB-2000}.
 However, the finite Coulomb gap above
 the $T_{c}$ requires the existence of tunnel barrier(s), presumably
unintentionally formed at the overlapping regions with bismuth contacts. The
observed Coulomb gap (Fig. 4a) can be simulated by a chain $\sim 10$ parallel-plate capacitors with the area $\sigma= 15 \times 15 ~ nm^{2}$ separated by a 1 nm vacuum
barrier. Note that formation of parallel junctions does not alter the Coulomb
gap \cite{Geigenmueler-Schoen-EPL-10-765-1989}. Hence, the major
part of the $20~ \mu m$ long high-Ohmic nanowires is metallic and
the QPS contribution should not be disregarded. If the same number of
junctions would be responsible for the Coulomb blockade and the Bloch
oscillations, then the values $\delta V_{n}$ for the steps 0/1
and 2/1 should be comparable \cite{Averin-Odintsov-FNT90-16},
which is not the case (Fig. 4b, inset). One may conjecture that several
serially connected junctions are responsible for the (large) Coulomb gap,
while smaller number of `active' elements -- for the Bloch steps. Those
serially connected junctions act as an additional high-impedance environment.
Given the equivalence of the quantum dynamics of a JJ and a QPSJ, our data
cannot distinguish whether that `active' element is a static JJ, or a driven
by quantum fluctuations dynamic QPSJ. However, whatever is the case, our
experiment is the clear evidence of Bloch oscillations. It has been suggested
that the proximity of a nanostructure material to superconductor-to-insulator
transition (SIT), actively studied in 2D systems \cite{Vinokur2007}, \cite{Vinokur2008},
\cite{Fegelman2010}, \cite{Sacepe2011}, \cite{Chand2012}, facilitate observation of the coherent QPS contribution
\cite{Astafiev-coherentQPS-NATURE2012}. Our present results on Ti
nanowires indicate that the SIT is not the mandatory requirement for
observation of the Coulomb phenomena in QPS-driven nanowires. The resistivity
of the samples is relatively low being on the metal side of the
metal-to-insulator transition.

The presence of the charge effects both in thicker, essentially metallic
structures (Figs. 2 and 3), and in the thinnest ones (Fig. 4), where several
tunnel junctions might have been unintentionally formed, supports the
universality of the phenomena originating from the fundamental duality of QPS
and Josephson systems. In addition to the importance of the discovery for
basic science, the observation of the Bloch singularities relating the current
$I_{n}$ and frequency $f_{RF}$ through the universal relation
$I_{n}=2e \times n \times f_{RF}$, can be
considered as the proof-of-principle demonstration of the qualitatively new
approach to the important metrological application -- the quantum standard of
electric current. Certainly, the demonstrated accuracy (Fig. 4c) is not yet
sufficient for practical metrology. However, the high absolute values of the
currents $I_{n}$ are very encouraging: they reach the nA range (Figs. 2
and 4) and by far exceed the alternative single electron solutions barely
providing $\sim 10^{-11} ~A$ currents \cite{Zimmerman-PhysToday2010}. 
The progress in the topic has the strong potential to revolutionalize the modern metrology.

\begin{acknowledgments}

The authors would like to acknowledge D.
Averin, D. Haviland, L. Kuzmin, Yu. Nazarov, A. Zaikin and A. Zorin for
valuable discussions and L. Leino for help with SPM analysis. The work was
supported by the Finnish Technical Academy project DEMAPP and Grant
2010-1.5-508-005-037 of Russian Ministry of Education and Research.

\end{acknowledgments}

\bibliography{PRL}

\end{document}